\documentclass[12pt]{article}

\usepackage{amsmath}
\usepackage[dvips]{epsfig}
\usepackage{amssymb}
\usepackage{color}
\usepackage{colordvi}
\usepackage{graphicx}
\usepackage{pdfsync}

\newtheorem{theorem}{Theorem}[section]
\newtheorem{lemma}[theorem]{Lemma}

\newtheorem{proposition}[theorem]{Proposition}

\begin{document}

\title{Maxwell's Equations, Hodge Theory,\\  and Gravitation}

\author{D.~H.~Sattinger\\
Department of Mathematics\\
University of Arizona\\
 Tucson, Arizona }

\date{October 10, 2013} 
 
\maketitle

\abstract{ A mathematical proof is given that Maxwell's equations are an {\it artifact} of Hodge theory together with the laws of Gauss and Amp\`ere, taken as axioms. They are thus geometric in nature, independent of any specific physical mechanisms, and valid for any force field -- attractive or repulsive -- generated by a material density and current. In particular, with appropriate sign changes to reflect the attractive nature of the field, they apply to gravitational fields on Minkowski space-time as well. 

The linearization of the Einstein Field Equations on Minkowksi space-time leads also to a linear theory of gravity; but this theory is spin 2, while Maxwell's field theory is spin 1. Hence the two theories are distinct. The relationship of Maxwell's field theory to Einstein's geometric theory is explained for weak fields. }
\medskip


\hskip 3mm

\noindent dsattinger@math.arizona.edu\\
\noindent http://math.arizona.edu/$\thicksim$dsattinger/

\medskip

\noindent 617 N. Santa Rita Ave.\\
\noindent P.O. Box 210089\\
\noindent Tucson, AZ 85721-0089 USA

\section{Introduction}\label{intro}

Maxwell obtained his equations of electrodynamics by assembling the results of the many physical experiments that had been carried out in the laboratory -- especially the Laws of Faraday and Amp\`ere -- and then introducing the Maxwell displacement $\bf D$. The gravitational field is so weak that it is impractical to carry out the corresponding experiments in the laboratory; but a mathematically rigorous derivation of Maxwell's equations for gravitation was recently given in \cite{dhs1}. The salient features of that derivation  were: i) a mathematical proof of Faraday's Law; and ii) a reversal of the orientation of Minkowski space-time $\mathfrak M^4$ to account for the attractive nature of gravity.  No physical experiment was cited to establish the physical validity of Maxwell's equations for gravitation; and the derivation required the hypothesis that $\mu\epsilon=c^{-2}$, where $\epsilon$ and $\mu$ are physical parameters associated with the laws of Gauss and Amp\`ere.

In the present paper, the proof of Faraday's Law is replaced by the use of the Hodge decomposition theorem; and the extension to attractive fields such as gravity is obtained by reversing the signs of  $\epsilon$ and $\mu$.  This approach is considerably simpler, and provides a unified treatment of force fields, both attractive and repulsive, on $\mathfrak M^4$ generated by a material density $\rho$ and current $\bf J$, independent of the particular physical mechanisms involved. The main result in the case of static fields is given in Theorem \ref{MaxHodge}. In Theorem \ref{equiv} we prove that $\mu\epsilon$ has the dimension of velocity to the power -2. Einstein's fundamental postulate of special relativity -- that the speed of light is a fundamental constant of nature -- then establishes the required hypothesis holds for general force fields. 

That Maxwell's equations can be written concisely in the language of differential forms, is well known (Bott; Flanders; Misner, Thorne and Wheeler; and Weinberg).  Bott explains the topological aspects of Hodge's work and discusses Maxwell's equations for the electromagnetic field and the Yang-Mills equations in connection with Hodge theory. However, his discussion is restricted to Maxwell's equations in a vacuum (no charge or current), and to the equations on $\mathbb E^4$, in which case they are elliptic.  
None of these authors show that Maxwell's equations can be extended to the gravitational field, and none of them make use of the Hodge decomposition.

The language of differential forms alone is not sufficient to prove that Maxwell's equations apply to gravity. Something additional is needed, such as a  proof of Faraday's Law in \cite{dhs1}, or  the Hodge  decomposition theorem as we do here.  Moreover, the laws of Gauss and Amp\`ere can be expressed in terms of Hodge duality, and this formulation is key to the extension of Maxwell's equations to attractive forces.

The Hodge theory of differential forms establishes the orthogonal decompositions 
$$\Lambda_1(\mathbb E^3)={\mathfrak E}\oplus {\mathfrak H}, \qquad \Lambda_2({\mathbb E^3}) =
{\mathfrak B}\oplus {\mathfrak D},$$
where $\mathfrak E$ and $\mathfrak H$ are  the subspaces of exact and co-exact 1-forms; while   
$\mathfrak B$ and $\mathfrak D$ are the subspaces of exact and  co-exact 2-forms.
 Given a stationary vector field ${\bf F}$ on $\mathbb E^3$, consider the associated differential forms $F={\bf F}\cdot d{\bf x}$ and  $\ast F= {\bf F}\cdot d{\bf S}$, where $d{\bf S}$ is an oriented surface element in $\mathbb E^3$. The Hodge decompositions show that $F$ and $\ast F$ are both sums of exact and co-exact differential forms. These four differential forms correspond to the four vector fields $\bf E,H$  and $\bf B, D$ of the electromagnetic field.   
 
 In \S\ref{statfields} it is shown that the two parameters $\epsilon$ and $\mu$ are determined by the postulates 
$$
\iiint\limits_U\rho\,dv =\epsilon\iint\limits_{\partial U} \ast F, \qquad  \oint\limits_\gamma F=\mu \iint\limits_{S_\gamma}{\bf J}\cdot d{\bf S},
$$
where $U$ is a smoothly bounded surface; $S_\gamma$ is a smooth surface spanning the closed curve $\gamma$;  $\rho$ is the material density, and ${\bf J}$  the material current.  These identities are equivalent to the laws of Gauss and Amp\`ere in the electromagnetic case; but otherwise the results are general and do not depend on the underlying physical mechanisms. The structure of Maxwell's equations is thus an artifact of Hodge theory and holds for general force fields generated by a material density $\rho$ and material current $\bf J$.

Maxwell's dynamical equations involve 2-forms on 4 dimensional Minkowski space-time $\mathfrak M^4$; and the system of partial differential equations $F=dA$ is then hyperbolic when $F$ is a 2-form. The relevant 2-forms are the Faraday 2-form $F$ and the Maxwell-Amp\`ere 2-form $G$; related to the laws of Faraday and Maxwell-Amp\`ere laws. These are Hodge duals of one another,  and the Lagrangian for Maxwell's equations is obtained in terms of them \eqref{Lagrangian},  \S\ref{dynamics}.

Efforts to extend Maxwell's field theory to gravitation date back to a paper published by Oliver Heaviside in 1893 \cite{OH} entitled ``A Gravitational and Electromagnetic Analogy''.  
Heaviside's attempt  was followed by Lorentz (1900) \cite{HAL} and Poincar\'e (1905) \cite{Poincare}. Poincar\'e believed that Lorentz invariance was a fundamental fact of physics; and at the end of his paper, in a section entitled {\it Hypoth\`eses sur la Gravitation}, he proposed a rudimentary  Lorentz-covariant form of the gravitational field in which {\it l'onde gravifique, $\dots$  \'etant suppos\'ee se propager avec la vitesse de la lumi\`ere $\dots$ peut se partager en trois composantes, la premi\`ere une vague analogie avec la force m\'ecanique due au champ \'electrique, les deux autres  avec la force m\'ecanique due au champ magn\'etique.}  

Those early attempts were abandoned with the success of Einstein's General Theory of Relativity. The historical arguments against a Lorentz invariant theory of gravity are laid out by Abraham Pais in Chapter 13 of his biography of Albert Einstein \cite{P}. Heaviside's paper is flatly dismissed; but those objections do not to apply to   {\it weak}  fields   \cite{dhs1}.

Theories of weak gravity are also obtained by linearizing the Einstein field equations at the Minkowski metric (see Landau and Lifshitz, \S107; Misner, Thorne and Wheeler, Chapter 18;  and Weinberg, Chapter 10). Maxwell's equations are a spin 1 theory, however, while the linearized Einstein equations are a spin 2 theory. This refers to the fact that the plane wave solutions of Maxwell's equations in a vacuum have helicity $\pm 1$ while those in Einstein's linearized equations have helicity $\pm 2$ (see the discussion in Weinberg). Thus the two theories of weak gravitation are distinct. 

This raises the question as to how Maxwell's field theory is related to Einstein's general theory of relativity in the case of weak gravity. Roughly speaking, on flat space-time, the metric tensor is given by the Minkowski metric, and the geodesic flows are straight lines. In that case, the Maxwell field dominates and determines the dynamics of mass systems at leading order; the energy-momentum tensor can be computed explicitly in terms of the Maxwell field and at next order the perturbation of the metric tensor determines the relativistic corrections. The perturbative method for weak fields, treated extensively in Weinberg, Chapter 9, is called the Post Newtonian approximation and will be sketched briefly in \S\ref{EM}.

\section{Hodge Theory}\label{helmhodge}

The Hodge decomposition is an extension of the classical Helmholtz decomposition which
asserts that any smooth vector field ${\bf F}$ on $\mathbb R^3$ can be decomposed as the sum of an irrotational and a solenoidal field,
\begin{equation}
{\bf F}=-\nabla \phi +\nabla \times {\bf A}.\nonumber
\end{equation}
Helmholtz was led to his theorem in his studies of fluid flow by attempting to resolve the velocity field into separate components of  irrotational (potential) and rotational (solenoidal) flows. In the theory of the electromagnetic field, $\phi$ and ${\bf A}$ are respectively the scalar and vector potentials. In the classical (Newtonian) theory of gravitation, however, it has always been presumed at the outset to be irrotational, despite the fact that there is not a shred of evidence, mathematical or physical, to support that assumption.  

Since  $\nabla\phi$ transforms as a vector field and $\nabla\times {\bf A}$ transforms as an anti-symmetric contravariant 2-tensor, their sum is not tensorial. This problem is resolved by reformulating the problem in terms of differential forms.  
To fix language and notation, we give here a quick review of the basic ideas.  We assume the reader is familiar with the basic operations of wedge product $\wedge$ and exterior derivative $d$ on $p$-forms $\Lambda_p$ and that the theorems of Green, Gauss, and Stokes are collected in a single theorem, known as Stokes' theorem,
\begin{equation}\label{stokes} \iint\limits_\Omega d\omega=\int\limits_{\partial \Omega} \omega.
\end{equation} Here, $\omega\in \Lambda_p$ has differentiable
coefficients, and $\Omega$ is a $p+1$ dimensional, oriented manifold embedded in
$\mathbb E^n$, with smooth boundary $\partial \Omega$. 

A  form $\omega\in \Lambda_p$ is said to be {\it closed} if $d\omega=0$ and {\it exact} if $\omega=d \chi$, where $\chi\in\Lambda_{p-1}$. Since $d^2=0$, a $p$-form is closed if it is exact. In a simply connected region the two conditions are equivalent; and it will be sufficient to restrict ourselves to this case. A necessary and sufficient condition for $\omega\in \Lambda_p$ to be exact in a region $U$ (not necessarily simply connected) is that its integral over every closed $p$ manifold $\Sigma\subset U$ vanish:
$$
\iint\limits_{\Sigma}\omega =0, \quad \text{whenever} \quad \partial\Sigma=\emptyset.
$$
In the special case of a 1-form $E$, $\Sigma$ is a closed path, and the integral above is a line integral called the {\it circulation}. If the circulation vanishes for every smooth closed path, then regardless of the topology of the region, $E$ is exact, and there exists a 0-form $\phi$ (that is, a single-valued function) such that $E=-d\phi$. It is standard convention to normalize the potential to vanish at infinity, so that it is  explicitly given by
\begin{equation}\label{potenergy}
\phi({\bf x})=\int_{\bf x}^\infty E, \qquad {\bf x}\in \mathbb E^3.
\end{equation}
In the static case the electric and gravitational fields are both conservative, so that their corresponding 1-forms  are exact; but the electrostatic potential is positive, while the gravitational potential is negative. This reflects the fact that the gravitational force is attractive, while the electrostatic force, defined in terms of  like charges, is repulsive.

Hodge's theory has evolved considerably since his book  was published, and some of the language and notation has changed. The book by Flanders, though not complete, nevertheless offers a readable introduction to the subject, especially for those interested in physical applications. 
The Hodge star operation maps $p$-forms to $n-p$-forms, where $n$ is the dimension of the manifold; it plays a fundamental role in potential theory, and is defined as follows: Given an oriented volume element $dv$ on a smooth $n$-dimensional manifold $\mathcal M$ and  $\omega\in \Lambda_p(\mathcal M)$, $\ast \omega$ is defined as the $n-p$ form for which $\omega\wedge \ast \omega=dv$. 
For example, the standard (right-handed) volume element on $\mathbb E^3$ is $dv=dx^1\wedge dx^2\wedge dx^3$. The associated Hodge star operation is 
\begin{equation}\label{*3}
\ast dx^i=dx^j\wedge dx^k, \qquad \ast 1=dv, \qquad \ast\ast=id.
\end{equation}
Here $i,j,k$ denote the integers 1,2,3 in cyclic order, and $id$ denotes the identity mapping. The star operation is extended to $\mathbb E^n$ in the obvious way; the reader may check that $\ast\ast=(-1)^{n+1}\ id \  {\rm on}\ \mathbb E^n.$

An inner product, called the Hodge duality, is defined for $\xi,\,\eta\in \Lambda_p(\mathbb E^n)$ by
\begin{equation}\label{Hdual}
(\xi,\eta)=\iiint\limits_{\mathbb E^n} \xi\wedge \ast \eta ,
\end{equation}
where $\ast$ is the operation associated with an oriented volume element $dv$. 
The Hodge duality implicitly defines a formal adjoint to the exterior derivative $d$.  Called the {\it coderivative}, it maps $\Lambda_{p+1}$ to $\Lambda_p$, and is defined by the relation 
\begin{equation}\label{adjoint}
(d\,\xi,\eta)=(\xi, \delta\eta) \qquad  \xi\in \Lambda_p,\ \eta\in\Lambda_{p+1}
 \end{equation} 
where $\xi,\,\eta$ have compact support on $\mathbb E^n$. On $\Lambda_p(\mathbb E^n)$ the co-derivative is given by $\delta=\delta_p =(-1)^{p-1}\ast^{-1}d\ast$  (\cite{dhs1}, Proposition 2.1). 

A differential form $\xi$ is {\it co-closed} if $\delta\xi=0$ and {\it co-exact} if $\xi=\delta \eta.$ The reader should check that $\delta\,:\,\Lambda_p\to\Lambda_{p-1}$, and that on a manifold with trivial topology such as $\mathbb E^n$,  $\delta^2=0$. A differential form is co-exact if and only if it is co-closed. A differential form is called harmonic if it is both closed and co-closed, i.e. $d\xi=\delta \xi=0$. On $\mathbb E^2$ these two equations are precisely the Cauchy-Riemann equations; and so this system of equations can be called the {\it generalized Cauchy-Riemann} equations. If $\xi$ is harmonic then $\Delta \xi=0$, where $\Delta=d\delta+\delta d$ is the Laplacian.

The Hodge decomposition on a compact manifold states that every (smooth) $p$-form $\omega$ can be decomposed as $\omega=d\xi+\delta \eta +\alpha$, where $\alpha$ is a harmonic form. Since $(d\xi,\,\delta\eta)=(\xi,\,\delta^2\eta)=0$, etc. it is clear that the subspaces of exact, co-exact, and harmonic forms are mutually orthogonal; hence the Hodge decomposition can also be written as
\begin{equation}\label{HodgeLp}
\Lambda_p=[\text{im}\,d]\oplus [\text{im}\,\delta]\oplus [\text{ker}\,d \cap \text{ker}\,\delta].
\end{equation}
This formulation of Hodge's theorem leads to a simple proof of its extension to $\Lambda_p(\mathbb E^n)$ based on the orthogonal decomposition theorem of Hilbert spaces:
\begin{proposition}\label{ODT} Let $\mathfrak H$ be a Hilbert Space, $\mathcal E$ a subspace and ${\mathcal B}={\mathcal E}^\perp$ its orthogonal complement. Then every element $F\in \mathfrak H$ can be uniquely decomposed as $F=E+B$, where $E\in\mathcal E$ and $B\in\mathcal B.$ 
We write $\mathfrak H=\mathcal E\oplus \mathcal B.$ \end{proposition}
 
We denote by $\Lambda_p(\mathbb E^n)$ the Hilbert space of $p$-forms with finite Hodge norms
$(F,F)<+\infty$ for $F\in \Lambda_p$. In the applications here, $n=3$ and the potentials, both scalar and vector, are solutions of \eqref{Poisson} and {\eqref{PoisA}  below  where $\rho$ has compact support. They are thus solutions of $\Delta \phi=0$, etc. in exterior domains; hence they decay as $r^{-1}$ as $r\to \infty$. The fields themselves are first order derivatives of the potentials, specifically, $\nabla\phi$ and $\nabla\times\bf A$. They therefore decay as $r^{-2}$ at infinity and are regular at the origin; hence they are square integrable.

To account for the lack of differentiability, the decomposition must incorporate the notion of weak derivatives. 
We say that $d\alpha =\beta$ in the weak sense if $(d\alpha,\eta)=(\beta,\delta\eta)$ for all differential forms $\eta\in \Lambda_{p+1}(\mathbb E^n)$ with $C^1$ coefficients and compact support. A similar definition applies to the equation $\delta \alpha=\beta$. Note that if  $d\alpha =\beta$ in the weak sense and $\alpha$ is itself $C^1$, then $\beta$ is continuous and $d\alpha=\beta$ in the ordinary (strong) sense.

\begin{theorem}\label{Hodgedecomp} The Hilbert spaces $\Lambda_p(\mathbb E^n)$ decompose into the direct sum of the $L^2$ exact and co-exact forms: $\Lambda_p=[d\Lambda_{p-1}]\oplus [\delta \Lambda_{p+1}]$, where $[d\Lambda_{p-1}]$ denotes the $L^2$ closure of the linear set $\{ dA\,:\,A\in \Lambda_{p-1}\}$, etc. and by default, $[d\Lambda_n]=[\delta \Lambda_0]=0$. Thus every differential form $F\in \Lambda_p$ can be written as $F=dA+\delta \Phi$, where $A\in \Lambda_{p-1}$ and $\Phi\in \Lambda_{p+1}.$

Moreover, since $\ast\ast=(-1)^{n+1}id$,
\begin{equation}\label{astdast}
\ast [d\Lambda_p]=[\delta \Lambda_{n-p}].
\end{equation}
\end{theorem}
{\sl Proof:}  The harmonic forms satisfy $d\alpha=\delta \alpha=0$, hence each term must be harmonic in the usual sense. By Liouville's theorem, each coefficient of $\alpha$ must therefore be a constant; and since the Hodge norm of $\alpha$ must be finite, those constants must vanish. So there are no harmonic components in the Hodge decomposition on Euclidean spaces. The orthogonal decomposition \eqref{HodgeLp} therefore follows immediately from Proposition \ref{ODT}.

The identity \eqref{astdast} follows very simply from the following:
$\ast [d\Lambda_p]= \{\ast d\alpha\,:\, \alpha\in \Lambda_p\} =\{\ast d\ast\, \beta\, : \, \beta=(-1)^{n+1}\ast\alpha\in \Lambda_{n-p}\}=
[\delta \Lambda_{n-p}]. \ \  \blacksquare $

\smallskip

Finally, we shall need the expression for $d{\bf S}$, introduced above, for the vector element of surface area on a 2 dimensional surface $S$ embedded in $\mathbb E^3$. It is given by $d{\bf S}=(X_u\times X_v)\, du\wedge dv$, where $X=(x^1(u,v),x^2(u,v),x^3(u,v))$ is a parametrization of a neighborhood of $S$ by local coordinates $u,v$. We leave it to the reader to verify the identities
$$
dx^i\wedge dx^j=\frac{\partial (x^i,x^j)}{\partial (u,v)} du\wedge dv,
\qquad B=B_j\,dx^k\wedge dx^l.
$$
Here and throughout this article, the expression for $B$ signifies a summation over $j,k,l$ from 1 to 3 in cyclical order.

\section{Stationary Field Theory}\label{statfields}

Maxwell's equations of electrodynamics in vector form are (see Stratton) 
\begin{alignat}{2}
& {\bf \nabla \times E}+\frac{\partial {\bf B}}{\partial t}=0, \qquad & {\rm div} \,{\bf B}=0; \label{Faraday}
\\[4mm]
&  {\bf \nabla \times H}-\frac{\partial {\bf D}}{\partial t}={\bf J},  & {\rm div}\, {\bf D}=\rho.
  \label{Ampere}
 \end{alignat}
 Here, $\bf E$ is the electric field, $\bf B$ the magnetic induction, $\bf H$ the magnetic field, and $\bf D$ the 
 electric displacement introduced by Maxwell. The first pair of equations constitute the differential form of Faraday's Law, and the second the Maxwell-Amp\`ere Law. The system is closed with the two constitutive relations 
 \begin{equation}\label{CR}
{\bf D}=\epsilon {\bf E}, \qquad {\bf B}=\mu {\bf H},
\end{equation}
where $\epsilon$ and $\mu$ are the electric and magnetic permittivities. 

Equations \eqref{CR} ignore the fundamentally different vectorial properties of $\bf E,\,H$ and $\bf D,\,B$.  Furthermore, the left side of \eqref{Ampere} is an axial vector, while the right side is a polar vector.  Once again, the difficulties are resolved by the use of differential forms.

Denote the Hodge decompositions of $\Lambda_1$ and $\Lambda_2$ on $\mathbb E^3$ by
$\Lambda_1={\mathcal E}\oplus{\mathcal H}$ and $ \Lambda_2={\mathcal B}\oplus {\mathcal D}$, where
$\mathcal E$ and $\mathcal B$ are respectively: ${\mathcal E}=[\{E\, :\,  E=-d\phi\}]$, $ {\mathcal B}=[\{B\,
: \, B=dA, \, A\in \Lambda_1\}]$.
By \eqref{astdast}, 
\begin{equation}\label{D*E}
{\mathcal D}= \ast {\mathcal E}, \qquad {\mathcal B}=\ast {\mathcal H}.
\end{equation} 
Let $\bf F$ be any square integrable stationary force field on $\mathbb E^3$; and let $F={\bf F}\cdot d{\bf x}$. By the Hodge decomposition 
\begin{equation}\label{HD1}
F=-d\phi+\ast dA,
\end{equation} where $\phi$ is a 0-form and $A$  a 1-form.

Consider the two integrals
$$
{\mathcal C}(\gamma)=\oint\limits_\gamma F, \qquad   {\mathcal F}(S)=\iint\limits_S \ast F
$$
where $\gamma$ is a smooth closed curve  and $S$ a smooth closed surface.   The line integral ${\mathcal C}(\gamma)$ is called the {\it circulation} of $F$ around $\gamma$; and ${\mathcal F}(S)$ is called the {\it flux} of  $F$ through $S$.  By \eqref{HD1},
\begin{equation}\label{integrals}
 {\mathcal F}(S) = -\iint\limits_S \ast d\phi, \qquad  {\mathcal C}=\oint \ast dA. 
 \end{equation}

Gauss' Law of electrostatics states that {\it the electric flux through a closed surface is proportional to the enclosed electric charge}, and is associated with the first integral above. It may be interpreted as a general axiom of conservative force fields, as follows. Let  ${\bf E}=-\nabla \phi$ be any conservative force field, and $E={\bf E}\cdot d{\bf x}=-d\phi$ the associated exact 1-form. 
Then 
$$
{\mathcal F}(S)= -\iint\limits_S \ast d\phi=\iint\limits_S \ast E=\iint\limits_S {\bf E}\cdot d{\bf S};
$$
hence $ {\mathcal F}(S)$ is precisely the flux of the lines of force through the closed surface $S$.

Let the material density be $\rho\ge 0$; since $\rho\,dv$ is a closed 3-form, there is a 2-form $D$ such that $dD=\rho\,dv.$ By Stokes' theorem, 
\begin{equation}\label{totalcharge}
{\mathcal Q}(S)=\iint\limits_S D= \iiint\limits_{U_S} \rho\,dv
\end{equation}
is the total amount of material contained within $S$.  
Gauss'  Law asserts that ${\mathcal Q}(S)= \epsilon {\mathcal F}(S)$ for all closed surfaces $S$; hence 
 \begin{equation} \label{GaussE}
 D =\epsilon\, \ast E, \qquad \text{Gauss' Law}.
\end{equation}
It follows  that 
\begin{equation}\label{deltaE}
\delta E=\ast d\ast E=\frac1\epsilon\ast dD=\frac\rho\epsilon.
\end{equation}

\begin{theorem} In the case of an inverse square law on $\mathbb E^3$,  $\epsilon$ is given by
\begin{equation}\label{epsg}
\epsilon=\frac1{4\pi G}, 
\end{equation}
where $G$ is the physical constant which determines the strength of the field  produced by $q$. The parameter $\epsilon$ is positive or negative, according as the field is repulsive or attractive.
\end{theorem}
{\sl Proof:} For a point source $q$ at the origin the displacement $D$ is given by
\begin{equation}\label{Dr2}
D=\frac{q}{4\pi}\frac{ x_j}{ r^3}\,dx^k\wedge dx^l.
\end{equation}
The reader may verify by direct calculation that $dD=0$ for $r>0$; hence the integral of $D$ over any closed surface $S$ enclosing the origin is equal to the integral over the sphere $S_R$ of radius $R$ centered at the origin. By Stokes' theorem
$$
\iint\limits_{S} D=\frac{q}{4\pi R^3}\iint\limits_{S_R} x_jdx^k\wedge dx^k=
\frac{q}{4\pi R^3}\iiint\limits_{B_R} 3\, dv =q,
$$
where $B_R$ denotes the interior of the sphere. 

 Let $E={\bf E}\cdot d{\bf x}$ be the 1-form associated with the field ${\bf E}$ produced by a point source $q$ at the origin. Then
\begin{equation}\label{Er2}
E=Gq \frac{{\bf \widehat r}\cdot d{\bf x}}{r^2}=Gq \frac{x_j dx^j}{r^3}; \qquad \ast E=Gq \frac{x_j}{r^3}\, dx^k\wedge dx^l.\end{equation}
Equation \eqref{epsg} follows by comparing \eqref{Dr2} and the second equation in \eqref{Er2}.

The lines of force exit or enter the region bounded by $S$, and ${\bf E}\cdot d{\bf S}$ is respectively positive or negative,  according as the force is repulsive or attractive.  Since $\rho\ge 0$ in either case, $\epsilon$ is positive or negative according as the force is repulsive or attractive.  $\blacksquare$

For the electrostatic field $G$ is called the Coulomb constant, after Coulomb, who first measured it in 1785 using a torsion balance he had developed. In the case of gravitation,  $G$ is called the Cavendish constant, after Henry Cavendish, who was the first to determine the constant accurately in 1798, also using a torsion balance. In the unified presentation here, the Coulomb constant is positive while the Cavendish constant is taken to be negative.

Now we turn to Amp\`ere's Law. In the stationary case, conservation of the material requires that $\nabla\cdot {\bf J}=0$, equivalently, that $d\ast J=0$. Consequently $\ast J={\bf J}\cdot d{\bf S}$ is a closed 2-form, and there is a 1-form $H$ such that 
 \begin{equation}\label{Ampslaw}
dH=\ast  J. 
\end{equation} In the language of vector analysis, \eqref{Ampslaw}  takes the form  $\nabla\times {\bf H}={\bf J}$, and is known as Amp\`ere's Law. The result, however, is an immediate mathematical consequence of the conservation of charge ($d\ast J=0$) and involves no physical assertion. For our purposes, it is more appropriate to link Amp\`ere's Law with the parameter $\mu$ and to frame it as a constitutive law, analogous to Gauss' law of electrostatics, as follows:  {\it the circulation of $F$ around a closed path $\gamma$ is proportional to the current passing through any surface spanning $\gamma$}:
\begin{equation}  
{\mathcal C}(\gamma)=
\mu \iint_{S_\gamma}  \ast J  \quad \text{Amp\`ere's Law}. \label{amp}
\end{equation}

\begin{theorem} Putting $B=dA$ in the Hodge decomposition \eqref{HD1}  we have
 \begin{equation}\label{BmuH}
 dB=0, \qquad  B=\mu \ast H \qquad \delta B=\mu J, \qquad \delta dA=\mu J
 \end{equation}
 \end{theorem}
 {\sl Proof:} By \eqref{integrals} and \eqref{Ampslaw} we have
$$
\iint_{S_\gamma}  \ast J=\iint_{S_\gamma}  dH=\oint\limits_\gamma H, 
\qquad {\mathcal C}(\gamma)=\oint\limits_\gamma \ast dA.
$$
Since \eqref{amp} holds for all closed loops $\gamma$, it follows  that $\ast dA=\mu H$, and  the relations \eqref{BmuH} follow immediately.  $\blacksquare$

Summarizing the preceding discussion, we have
\begin{theorem}\label{MaxHodge} The fields $\bf E, B, H,D$ of electromagnetic theory arise naturally in the Hodge decompositions of $\Lambda_p(\mathbb E^3)$ for $p=1,2$. The 1-forms $E={\bf E}\cdot d{\bf x}$ and $H={\bf H}\cdot d{\bf x}$ are respectively the exact and co-exact 1-forms; while $B={\bf B}\cdot d{\bf S}$ and $D={\bf D}\cdot d{\bf S}$ are the exact and co-exact 2-forms. The Laws of Gauss and Amp\`ere take the form $D=\epsilon\ast E$ and $B=\mu\ast H$, where $\epsilon$ and $\mu$ are the electric and magnetic inductances, and $\ast$ is the Hodge star operation on $\mathbb E^3$. The physical parameters $\epsilon$ and $\mu$ are positive or negative, according as the force is repulsive or attractive. \end{theorem} 

For future reference, we derive the equations for $\phi$ and ${\bf A}$ in vector notation. For the irrotational component ${\bf E}$ we have ${\bf E}=-\nabla \phi,$ $ {\bf D}=\epsilon {\bf E}$, $\nabla\cdot {\bf D}=\rho$, hence
\begin{equation}\label{Poisson}
\Delta\phi=-\frac\rho\epsilon.
\end{equation}
This is Poisson's equation for the potential of the irrotational component of the field.  For the vector potential we have
${\bf B}=\nabla \times {\bf A}$, $\nabla\times {\bf B}=\mu {\bf J}$, hence 
\begin{equation}\label{PoisA}
\nabla\times\nabla\times {\bf A}=\mu {\bf J}.
\end{equation}
Without loss of generality, we can assume that ${\rm div}\cdot{\bf A}=0$; hence, by the vector identity
$\nabla\times\nabla\times {\bf a}=\nabla(\nabla\cdot {\bf a})-\Delta {\bf a}$,
equation \eqref{PoisA} reduces to
\begin{equation}\label{DeltaA}
\Delta {\bf A}=-\mu {\bf J}.
\end{equation}

\section{Units and Dimensions}\label{units}

The derivations of  the static equations in the previous section were purely mathematical, but their application to physical cases requires that we attach physical dimensions to the quantities of interest. 
In particular, Maxwell's equations of electrodynamics require for their validity that 
\begin{equation}\label{sol1}\frac1{\epsilon\mu}=c^2,\end{equation}
where $c$ is the speed of light. The proof of \eqref{sol1} does not follow from the static equations alone; it requires the dynamical equations, as well as basic results from special relativity, including Einstein's axiom of special relativity that the speed of light is a fundamental constant of nature. In this and the next section we shall establish \eqref{sol1} for general force fields, both attractive and repulsive.

The relationship \eqref{sol1} was already known to Maxwell for the electromagnetic field, and he cited it specifically in his paper. For the case of gravitation, only the units of mass, length, and time, denoted by $m,\ell$, and $\tau$, enter the discussion; but when the field is generated by a material other than mass, an additional unit is needed. For electromagnetism, that unit is charge. In general, we shall assume that the force field is generated by a material density $\rho$ called charge and measured in units denoted by $q$. Thus, the basic physical units of measurement of a general force field are mass $m$, length $\ell$, time $\tau$, and charge $q$.
In the special case of gravitation, the material is mass, and $q=m$. 

The primary variables of dynamics are velocity $v$, acceleration $a$, force $f$, momentum $p$, and energy $\varepsilon$ with physical dimensions 
$$
v=\frac\ell\tau,\ \ a=\frac\ell{\tau^2}, \ \ f=ma, \ \ p=mv, \ \ \varepsilon=mv^2.
$$
The material density $\rho$ has dimensions $q\slash \ell^3$.

The force field ${\bf E}$, has dimensions of force per unit charge, hence 
\begin{equation}\label{[E]}
 [{\bf E}]= \frac{m}{q}\frac{\ell}{\tau^{2}} , \qquad [\ast E]=[{\bf E}\cdot d{\bf S}]=\frac{m}q \frac{\ell^3}{\tau^2}.
\end{equation}
The displacement 2-form $D={\bf D}\cdot d{\bf S}$ must have the dimension of $q$, since its integral over a 2 dimensional surface $S$ produces the total charge contained within; hence it follows by Gauss' Law \eqref{GaussE} that
\begin{equation}\label{[epsilon]}
[\epsilon]=\frac{q^2}m \frac{\tau^2}{\ell^3}.
\end{equation}

The magnetic field $\bf H$ is generated by the current $\bf J$. We have
$$
[{\bf J}]= [\rho {\bf v}]=\frac{q}{\ell^3}\frac\ell\tau=\frac{q}{\ell^2\tau}, \qquad [\ast J]=[{\bf J}\cdot d{\bf S}]=\frac{q}\tau.
$$
Now note that the exterior derivative $d$ is homogeneous of degree 0 with respect to dimension; that is
$[d\omega]=[\omega]$ for any p-form $\Omega$. It therefore follows from \eqref{Ampslaw} that $[H]=q\slash \tau$; and, since $[H]=[{\bf H}]\cdot \ell$,
\begin{equation}\label{[H]}
[{\bf H}]=\frac{q}{\ell \tau}.
\end{equation}
Equations  \eqref{[E]}, \eqref{[epsilon]}, and \eqref{[H]} are in agreement with  Stratton, \S1.8.

The units of $B={\bf B}\cdot d{\bf S}$ in the case of the electromagnetic field are defined to be {\it webers}; but to express webers in terms of the basic units, Stratton uses Faraday's Law to obtain 
\begin{equation}\label{[B]}
[B]=[{\bf B}\cdot d{\bf S}]=1\  weber=1\  \frac{kilogram \cdot meter^2}{coulomb\cdot second}.
\end{equation}
Since $[B]=[{\bf B}]\ell^2$, two immediate corollaries of \eqref{[B]} are
\begin{equation}\label{[B]v}
[{\bf B}]=\frac{m}{q\tau} \quad {\rm and}\quad   [{\bf B}][{\bf v}]={[\bf E}],
\end{equation}
where ${\bf v}$ is a velocity. Note that these two statements  are equivalent.

The second equation above implies that $\bf [v\times B]=[E]$, hence that $q({\bf E}+\,{\bf v}\times {\bf B})$ has the dimensions of force. This expression is of course known as the Lorentz force in electromagnetism. The following theorem shows that the same result holds for any force field, including the gravitational field.
\begin{theorem}\label{equiv}
The two statements in  \eqref{[B]v} are equivalent to the statement that  $[\mu\epsilon]=\tau^2\ell^{-2}$.
The relation $ [{\bf B}] c={[\bf E}]$ is a consequence of  Lorentz invariance. 
\end{theorem}

{\sl Proof:} If  $[\mu\epsilon]=\tau^2\ell^{-2}$, then by \eqref{[epsilon]} we find that $[\mu]=m\ell\slash  q^2$. By \eqref{BmuH} and \eqref{[H]} we have
$$
[B]=[\mu\ast H]=[\mu][{\bf H}\cdot d{\bf S}]=\frac{m\ell^3}{q^2}\frac{q}{\ell \tau}=\frac{m}{q}\frac{\ell^2}\tau,
$$
and the first statement in  \eqref{[B]v} follows. The argument is reversible. 

The proof of the second statement follows from  \eqref{BE} below, and the fact that all components of the potential $A_j$ have the same dimension. 

\section{Dynamical field theory}\label{dynamics}

Hodge theory for manifolds embedded in $\mathbb E^n$ leads to elliptic systems of partial differential equations; whereas Maxwell's dynamical equations for the electromagnetic field 
are hyperbolic. They are the field theory of special relativity, and are 
are naturally formulated on Minkowski space-time $\mathfrak M^4$. This space is obtained from $\mathbb E^4$ by setting $x_4=ict$. This is the point of view taken in Stratton, and we shall follow it here. 

Substantive differences arise when one attempts to extend Hodge's formalism to $\mathfrak M^4$. The equation $F=dA$ where $F$ is a 2-form, for example, is now a hyperbolic system. (See the proof of Faraday's Law in \cite{dhs1}.)  The Hodge duality on $\Lambda_2(\mathfrak M^4)$ is indefinite, so that the space is no longer a Hilbert space; and the proof of the Hodge decomposition given in \S\ref{helmhodge} is not valid. The discussion of Maxwell's equations and Yang Mills theory in Bott's article is restricted to the Euclidean space $\mathbb E^4$, since the Hodge decomposition of 2-forms is explicitly limited to the elliptic case. The situation can be finessed to a certain degree, however; and the Hodge decomposition on $\mathfrak M^4$ is not needed. Nevertheless, Maxwell's dynamical equations arise naturally out of this extended formalism, and are obtained  for both repulsive and attractive forces.  

The Hodge star operation on $\mathbb E^4$ associated with the oriented volume element $dv\wedge dx^4$ is
\begin{alignat}{2}
&\ast dx^j= dx^k\wedge dx^l\wedge dx^4 \qquad &\ast\, dx^4=-dv \label{1form}\\[4mm]
&\ast dx^j\wedge dx^k= dx^l \wedge dx^4  &\ast \,  dx^j\wedge  dx^4=dx^k\wedge dx^l  \label{2form}\\[4mm]
&\ast dv=dx^4,  &\ast\,  dx^j\wedge dx^k\wedge dx^4=-dx^l.\label{3form}
\end{alignat}
Note that $\ast\ast=-id$ in 4 dimensions. 

The star operation on $\mathfrak M^4$ is obtained by simply setting $x_4=ict$ in these equations.
The Hodge duality on $\mathfrak M^4$ is given by
\begin{equation}\label{HD4}
(A,B)=\frac1{ic}\int\limits_{\mathfrak M^r} A\wedge\ast B.
\end{equation}
Using this, the reader may verify that the Hodge duality is positive definite on $\Lambda_p(\mathfrak M^4)$ for $p\ne 2$. 

\begin{lemma}\label{Fstructure} Every exact form in $\Lambda_2(\mathfrak M^4)$ can be written as 
\begin{equation}\label{F2}
  F= E\wedge dt +B,
  \end{equation}
  where $ E=E_jdx^j,$ and $ B=B_i dx^j\wedge dx^k$, the sums running over $1\le i,j,k\le 3,$  is an exact 2-form.
  \end{lemma}
 {\sl Proof:}  The exact forms in $\Lambda_2$ are given by $F=dA$, where $A=A_jdx^j$, and 
    \begin{align*}
   dA=&
   \sum_{j<k\le 3}\left(\frac{\partial A_k}{\partial x^j}-\frac{\partial A_j}{\partial x^k}\right) dx^j\wedge dx^k\\
 &  \hskip.75in +\sum_{j=1}^3 \left(\frac{\partial A_4}{\partial x^j}-\frac{\partial A_j}{\partial x^4}\right) dx^j\wedge dx^4.
   \end{align*}
Putting 
  \begin{equation}\label{BE}
    B_i=\frac{\partial A_k}{\partial x^j}-\frac{\partial A_j}{\partial x^k},
  \qquad
 \frac{E_j} {ic}=\left(\frac{\partial A_4}{\partial x^j}-\frac{\partial A_j}{\partial x^4}\right). 
  \end{equation}
  we obtain \eqref{F2}.  $\blacksquare$ 
  
\smallskip

 Since the components of the Lorentz transformations are dimensionless variables, all coefficients $A_j$ of the 4-potential $A=A_jdx^j$ have the same dimension. It follows from \eqref{BE} that $[{\bf E}] = [{\bf B}] c,$ which is the second statement in Theorem \ref{equiv}.

\smallskip
  
  $F$ is called the Faraday 2-form.  Note that if we put $A_4=-\phi\slash ic$ we obtain
  $$
  E_j=-\frac{\partial \phi}{\partial x^j}-\frac{\partial A_j}{\partial t}, \quad \text{or} \quad {\bf E}=-\nabla\phi -\frac{\partial {\bf A}}{\partial t},
  $$ 
  which is the standard representation of the electric field in terms of the scalar and vector potentials in electromagnetic theory (see Stratton, \S1.21).

    \begin{lemma}\label{*F} The Hodge star operation interchanges exact $p$ forms with co-exact $4-p$ forms; and $\lambda$ is a real form if and only if  $\ast \lambda$ is imaginary.
  \end{lemma}
  {\sl Proof:} If $\lambda$ is exact, then $\lambda=d\alpha$, and $\gamma=\ast \lambda=\ast d \alpha=\delta \beta$, where $\beta=\ast \alpha$. Conversely,
  if $\gamma$ is a co-exact, then $\gamma=\delta \beta=\ast d\alpha$,  where $\alpha=\ast\,\beta$. Hence $\ast \gamma$ is exact. 
 
We shall say that a basis form on $\mathfrak M^4$ is time-like if it contains $dx^4$  and space-like if it does not. A $p$-form $\lambda$ is real if and only if $\overline \lambda=\lambda$ and imaginary if $\overline \lambda=-\lambda$. 
Therefore a differential form is real if and only if all its time-like coefficients are imaginary while all its space-like coefficients are real; and it is imaginary if the opposite holds. Since the Hodge star operation interchanges  space-like and time-like basis forms, it interchanges real and imaginary differential forms. $ \blacksquare$

In the static case we have $\ast B=\mu H$, and $\ast D=\epsilon E$ where $\ast$ is the Hodge operation on $\mathbb E^3$. The forms $D$ and $H$ and the parameters $\epsilon$ and $\mu$ were defined in terms of the Hodge decompositions on $\Lambda_1$ and $\Lambda_2$ over $\mathbb E^3$, together with the Laws of Gauss and Amp\`ere. When $\ast$ is the Hodge operation on $\mathfrak M^4$ we formally obtain 
  \begin{align}
    \ast B=&\ast B_jdx^k\wedge dx^l=\mu H_jdx^j\wedge dx^4=ic\mu H\wedge dt; \label{H}\\[4mm]
  \ast E\wedge dt=&\ast (\frac{E_j}{ic} dx^j\wedge dx^4) \nonumber \\[4mm]
   =&\frac{E_j}{ic} dx^k\wedge dx^l =\frac1{ic\epsilon} D_jdx^k\wedge dx^l=\frac1{ic\epsilon} D \label{D}
  \end{align}
  These two relations formally define $H$ and $D$ in the dynamic case and provide the dynamic version of the laws of Amp\'ere and Gauss.

 In Special Relativity, the material density $\rho$ and current ${\bf J}=\rho {\bf v}$ are combined into a single source term, the 1-form $J =J_jdx^j+ic\rho dx^4$. The reader may verify that $(J,J)>0$. Conservation of material (charge or mass) is expressed by the condition $\delta J=0$. The 4-potential is given by the 1-form $A=A_jdx^j.$  By the Hodge decomposition of $\Lambda_1(\mathfrak M^4)$, we may write $A=d\phi+\delta \Phi$, but since the field is obtained from $dA$, the term $d\phi$ has no effect; hence we may assume that $\delta A=0$. 

 \begin{theorem}\label{FarAmp} Every co-exact form $G\in \Lambda_2(\mathfrak M^4)$ can be written as $G=ic(H\wedge dt-D)$, where $H$ and $D$ are defined by equations \eqref{H} and \eqref{D}. $G$ satisfies the system
 \begin{equation}\label{MaxAmp}
\delta G=0, \qquad § dG=\ast J, 
 \end{equation}
 where $J$ is the current. If $F$ is a Faraday 2-form, then  $\ast F=\mu G$; and $F$ satisfies the system of equations
\begin{equation}\label{FarLaw}
dF=0, \qquad \delta F= \mu J.
\end{equation} 
Putting $F=dA$ ,  we have $\delta d A=-\square A=\mu J$, where 
\begin{equation}\label{squareA}
\square A=\sum_{j=1}^4 \square A_jdx^j, \qquad \square =\sum_{j=1}^3\frac{\partial^2 }{\partial x_j^2}-\frac1{c^2}\frac{\partial^2 }{\partial t^2}.
\end{equation}
 \end{theorem}
  {\sl Proof:} We begin by noting that  $\ast\ast=(-1)^p id$ and $\delta =(-1)^{p-1}$ on $\mathbb E^4$; hence $\delta=-\ast d\ast$ on  $\mathfrak M^4$.
    
Since $\delta J=d\ast J=0$, $\ast J$ is a closed 3-form, and the second equation in \eqref{MaxAmp} is solvable. Observe that  $G$, with $H=H_jdx^j$ and $D=D_j\,dx^k\wedge dx^l$, represents a general 2-form on $\mathfrak M^4$. By direct calculation
\begin{equation}
dG= \left(\frac{\partial H_l}{\partial x^k}-\frac{\partial H_k}{\partial x^l}-ic \frac{\partial D_j}{\partial x^4}\right)dx^k\wedge dx^l\wedge dx^4-ic\sum_{j=1}^3 \frac{\partial D_j}{\partial x^j}\,dv   \label{dG}
\end{equation}

Since $\ast J= (J_j\,dx^k\wedge dx^l\wedge dx^4 -ic\rho\,dv)$,   \eqref{MaxAmp}  and \eqref{dG} give the Maxwell-Amp\`ere equations \eqref{Ampere}:
$$
\left(\frac{\partial H_l}{\partial x^k}-\frac{\partial H_k}{\partial x^l}-ic \frac{\partial D_j}{\partial x^4}\right)=J_j, \qquad
\sum_{j=1}^3 \frac{\partial D_j}{\partial x^j}=\rho.
$$

By \eqref{H} and \eqref{D}, Lemma \ref{Fstructure}, and \eqref{sol1} we obtain
   \begin{align}\ast\,F=&\ast \left(\frac{E}{ic}\wedge dx^4+B\right)=\frac1{ic\epsilon}D
+\mu H\wedge dx^4 \nonumber \\[4mm]=&\mu\left(H\wedge dx^4+\frac1{ic\epsilon\mu}D\right) =ic\mu (H\wedge dt-D)
\label{*F=muG}
 \end{align}
 We call the 2-forms $G=ic(H\wedge dt-D)$ the Maxwell-Amp\`ere forms.  
  
 Since $J$ is a 1-form and $F,G$ are 2-forms, 
 $$
 \delta F=-\ast d\ast F= -\mu \ast dG=-\mu\ast\ast J=\mu J,
 $$
which is the second equation in \eqref{FarLaw}. It follows that $\delta d A=\delta F=\mu J$. 

Finally, let us show prove \eqref{squareA} when $A$ satisfies the Lorentz gauge.
Denoting $\partial_k A_j$ by $A_{j,k},\ etc.$ we have
\begin{align*}
dA=&\sum_{i,j=1}^4 A_{i,j}dx^j\wedge dx^i=\sum_{i,j=1}^4 A_{j,i}dx^i\wedge dx^j, \\
\ast dA=& \sum_{i,j=1}^4 A_{j,i}dx^k\wedge dx^l, \qquad k,l\ne i,j
\end{align*}
where $dx^i\wedge dx^j\wedge dx^k\wedge dx^l$ is equal to the standard volume element $dv\wedge dx^4$.
Then
$$
d\ast dA=\sum_{i,j,m=1}^4 A_{j,im}dx^m\wedge dx^k\wedge dx^l.
$$
Since $dx^m\wedge dx^k\wedge dx^l=0$ unless $m=i$ or $j$, this sum reduces to
$$
\sum_{i,j=1}^4 A_{j,ii}dx^i\wedge dx^k\wedge dx^l+\sum_{i,j=1}^4 A_{j,ij} dx^j\wedge dx^k\wedge dx^l,
$$
and
$$
\ast d\ast d A= -\sum_{i,j=1}^4 A_{j,ii} dx^j+\sum_{i,j=1}^4 A_{j,ij} dx^i.
$$
Now the first sum is $-\sum_j\Delta A_j dx^j$, while the second sum is 
$\sum_i \partial_i(\sum_j \partial_j A_j)dx^i=0$ by the Lorentz condition.
The Laplacian is $\Delta =\sum_j \partial_j^2$ on $\mathbb E^4$; but  on $\mathfrak M^4$
 it becomes the wave operator with the substitution $x_4=ict$. This completes the proof of Theorem \ref{FarAmp}.

From Theorem \ref{FarAmp} and  \cite{dhs1} \S6 we obtain the Lagrangian for Maxwell's equations in the form
\begin{align}
L\,dv\wedge dt = & \frac1{ic} Ldv\wedge dx^4= \wedge\frac1{ic}\left[\frac12 F\wedge\ast F+\mu A\wedge\ast J\right]    \nonumber\\
=&\frac\mu{ic}\left[\frac12 F\wedge G+A\wedge\ast J\right] .\label{Lagrangian}
\end{align}

\section{The Post Newtonian Approximation}\label{EM}

General Relativity posits two tensors, the metric  tensor and the energy momentum tensor, coupled by the Einstein Field Equations
\begin{equation}\label{EFE}
R_{\mu\nu}-\frac12 R\, g_{\mu\nu}=\frac{8\pi G}{c^4}T_{\mu\nu}
\end{equation}
where $R_{\mu\nu}$ is the Ricci curvature tensor, $R$ the scalar Ricci curvature, $g_{\mu\nu}$ the metric tensor,  $G$ the gravitational constant (taken here to be positive according to standard convention), $c$ the speed of light, and $T_{\mu\nu}$ the energy-momentum tensor. 
The parameter multiplying the energy-momentum tensor on the right-hand side is extremely small:
\begin{equation}\label{Rparam} 
\frac{8\pi G}{c^4}=2.068 \times 10^{-43}\frac{ sec^2}{ kg\,m}.
\end{equation}
Consequently, the right-hand side of \eqref{EFE} is nil except where the energy-momentum tensor is extremely large, such as in the vicinity of a massive star. 

If the right hand side is set to zero, we obtain the Einstein equations
$$
R_{\mu\nu}-\frac12 R\, g_{\mu\nu}=0.
$$
The Minkowski metric tensor is a trivial solution of these equations; but there are non-trivial solutions,
the Schwarzschild and Kerr metrics for example, as well. Non-trivial solutions of the Einstein equations have singularities, known as event horizons, at a finite distance from the source. They are normalized to tend to the Minkowksi metric at infinity. Thus  space-time is essentially flat far from mass sources, and the geodesic flow due to the metric tensor consists of straight lines. In such regions the Maxwell field dominates and determines the dynamics of mass distributions.

In the case of a Maxwell field, $T_{\mu\nu}$  can be computed explicitly from the Faraday 2-form (Landau and Lifshitz \S33), hence directly in terms of the field variables for either the electromagnetic or gravitational case. The calculation in \cite{LL} is carried out relative to the line element $ds^2=c^2dt^2-(dx^2+dy^2+dz^2)$, for which metric tensor is the diagonal matrix $(1,-1,-1,-1)$. In order to conform to their notation, as well as to the usual notation of General Relativity, we replace $x^4=ict$ with $x^0=ct$. The Faraday 2-form is then
$F=E\wedge dt+B=-dx^0\wedge c^{-1}E+B=F_{jk}dx^j\wedge dx^k$, with
$$
F_{jk}=\begin{pmatrix} 0 & - \displaystyle \frac{E_1}{c}&  -\displaystyle\frac{E_2}{c}&  -\displaystyle\frac{E_3}{c} 
\\[4mm]
\displaystyle \frac{E_1}{c }& 0 & B_3 & - B_2   \\[4mm]
\displaystyle\frac{E_2}{c}& -B_3 &  0 & B_1  \\[2mm]
\displaystyle\frac{E_3}{c} &  B_2 & -B_1 & 0 \end{pmatrix}.
 $$
 The energy-momentum tensor is then obtained from 
\begin{equation}\label{emt}
 T^{jk}=\frac1{4\pi}\left( -F^{i\ell}F_{\ell^k} +\frac14g^{jk}F_{\ell m}F^{\ell m}\right)
 \end{equation}
 
For $1\le j,k\le 3$  We find
 \begin{equation}\label{sigik}
 T^{jk}=\frac\mu{8\pi} \left(\delta_{jk} (\epsilon E^2 +\mu H^2)-2(\epsilon E_j E_k+\mu H_jH_k) \right);
 \end{equation}
 while for row 0 we obtain
 \begin{equation}\label{T0}
 T^{00}= \frac\mu{8\pi} (\epsilon E^2+\mu H^2), \qquad T^{0j}=\frac\mu{4\pi}\frac{E_kH_\ell-H_kE_\ell}{c} ,                   \qquad\end{equation} where $ j,k,\ell$ are the integers 1 through 3 in cyclic order.

 The  $3\times 3$ tensor $T_{ik}=-T^{ik}=\sigma_{ik},\ 1\le j,k\le 3$ is known as the Maxwell stress tensor. All components of the energy-momentum tensor have the physical dimensions of energy density. One of the classical objections to Maxwell's equations for gravitation was that the energy of the field was negative; but \eqref{T0} shows that the energy density $T^{00}$ is positive for {\it both} the electromagnetic and gravitational cases, since $\epsilon$ and $\mu$ are of the same sign in both cases. 

One of the first successes of General Relativity was Einstein's calculation of the precession of the perihelion of Mercury by solving his equations for a point mass in a vacuum. The energy-momentum tensor was assumed to vanish for all $r>0$. Thorne (\cite{Thorne}, p. 95) points out that the total advance of the perihelion is 1.38 seconds of arc per revolution, 1.28 of which are accounted for by Newtonian theory. Thus the solution based on classical, Newtonian mechanics is highly accurate, and Einstein's calculation can be regarded as the ``relativistic correction''. The Schwarzschild radius of the Sun is approximately 3 kilometers, but the effects of General Relativity on the planet Mercury are still observable. On Earth, however, these effects are no longer noticeable, and space-time in the vicinity of the Earth is virtually flat.

The fully relativistic flows about a central force field are generally modeled as the geodesic flow under the Schwarzschild metric.  In general, however, it is not possible explicitly to calculate the metric tensor associated with a solution of Einstein's equations. The perturbation scheme of computing higher order approximations is called the Post-Newtonian approximation, and has been developed extensively; see, for example, Weinberg, Chapter 9, and  Misner, Thorne, and Wheeler, Chapter 39. Maxwell's field theory constitutes the leading approximation in that scheme.   

Weinberg points out that ``the Newtonian effects of the planets' gravitational fields are of an order of magnitude greater than the first corrections due to General Relativity, and completely swamp the higher corrections that are in principle provided by the exact Schwarzschild solution.'' Thus, we should expect in general that the Maxwell field is the dominant component of the force field, with the equations of General Relativity supplying only a small correction, as they do in the case of Mercury.

  \end{document}